# *Quantum Computers and Quantum Computer Languages:*

# *Quantum Assembly Language and Quantum C[*]*

### By

### Stephen Blaha[**]

---

# ABSTRACT

We show a representation of Quantum Computers defines Quantum Turing Machines with associated Quantum Grammars. We then create examples of Quantum Grammars. Lastly we develop an algebraic approach to high level Quantum Languages using Quantum Assembly language and Quantum C language as examples.

**COPYRIGHT NOTICES AND ACKNOWLEDGMENTS**

# CONTENTS

# 1

# Probabilistic Computer Grammars

**Probabilistic Computer Grammars**
The preceding chapter described the production rules for a deterministic grammar. The left side of each production rule has one, and only one, possible transition.

Non-deterministic grammars allow two or more grammar rules to have the same left side and different right sides. For example,

$$A \to y$$
$$A \to x$$

could both appear in a non-deterministic grammar.

Non-deterministic grammars are naturally associated with probabilities. The probabilities can be classical probabilities or quantum probabilities. An example of a simple non-deterministic grammar is:

The head symbol is the letter S. The terminal symbols are the letters x and y. The production rules are:

| | | |
|---|---|---|
| S → xy | Rule I | |
| x → xx | Rule II | Relative Probability = .75 |
| x → xy | Rule III | Relative Probability = .25 |
| y → yy | Rule IV | |

The probability of generating the string xxy vs. the probability of generating the string xyy from the string xy is

xy → xxy    relative probability = .75

vs.

xy → xyy    relative probability = .25

The string xxy is three times more likely to be produced than the string xyy.

For each starting string one can obtain the relative probabilities that various possible output strings will be produced.

A more practical example of a Probabilistic Grammar can be abstracted from flipping coins – heads or tails occur with equal probability – 50-50. From this observation we can create a little Probabilistic Grammar for the case of flipping two coins. Let us let h represent heads and t represent tails. Then let us choose the grammar:

| | |
|---|---|
| S → hh | |
| S → tt | |
| S → ht | |
| S → th | |
| h → t | probability = .5 (50%) |
| h → h | probability = .5 (50%) |
| t → h | probability = .5 (50%) |

$$t \rightarrow t \qquad \text{probability} = .5 \ (50\%)$$

The last four rules above embody the statement that flipping a coin yields heads or tails with equal probability (50% or .5).

Now let us consider starting with two heads hh. The possible outcomes and their probabilities are:

$$hh \rightarrow hh \qquad \text{probability} = .5 * .5 = .25$$
$$hh \rightarrow th \qquad \text{probability} = .5 * .5 = .25$$
$$hh \rightarrow ht \qquad \text{probability} = .5 * .5 = .25$$
$$hh \rightarrow tt \qquad \text{probability} = .5 * .5 = .25$$

If we don't care about the order of the output heads and tails then the probability of two heads hh → ht or th is .25 + .25 = .5.

This simple example shows the basic thought process of a non-deterministic grammar with associated probabilities.

The combination of a non-deterministic grammar and an associated set of probabilities for transitions can be called a *Probabilistic Grammar*. We will see that the grammar production rules for the Standard Model must be viewed as constituting a Probabilistic Grammar™ with one difference. The "square roots" of probabilities – probability amplitudes are specified for the transitions in the grammar. Probability amplitudes are required by the Standard Model because it is a quantum theory. Therefore we will call the grammar of the Standard Model a *Quantum Probabilistic Grammar™*.

**Quantum Probabilistic Grammar**
Physics examples are presented here in the book. There are additional chapters also omitted.

# 2

# Quantum Turing Machines

## What are Turing Machines?
The linguistic view of the Standard Model leads to a number of questions. One important question is the nature of the Turing machine that accepts this language. A Turing machine is a generalized theoretical computer that is often used to analyze computational questions in computer science.

A personal computer can be viewed as a special purpose Turing machine. A personal computer has memory in the form of RAM and hard disks. A personal computer has built-in programs that tell it what to do when data is input into the computer. Similarly a Turing machine also has memory and instructions within it telling it how to handle input and how to produce output from a given input.

When we type input on a computer keyboard or have input come from another source such as the Internet or a data file, the input has to be in a form that the computer can handle. Similarly, the input for a Turing machine must have a specific form for the Turing machine to accept it, process it and then produce output. In the case of a Turing machine we say the input must be presented in a language that the Turing machine "accepts". In this context the word "accepts" means a format that the Turing machine can recognize and analyze so that it can process the data to produce output.

The language of the Standard Model is a type 0 computer language. A type 0 language requires a Turing machine to handle its productions. Because particle transitions are quantum and because the left side of a production rule can have several possible right sides (For example, a photon can transition to an electron-positron pair, a quark-antiquark pair and so on.) the Turing machine for the Standard Model language must be a non-deterministic Quantum Turing Machine.

**Features of Normal Turing Machines**
Before examining a Quantum Turing Machine for the Standard Model we will look at the features of "normal" Turing machines. A normal Turing machine consists of a finitely describable black box (its features are describable in a finite number of statements) and an infinite tape. The tape plays the role of computer memory. The tape is divided into squares. Each square contains a symbol or character. The character can be the "blank" character or a symbol. A tape contains blank characters followed by a finite string of input symbols followed by blank characters.

The black box consists of a control part and a tape head. The control part has a finite set of rules built into it (the "program") and a finite memory that it uses as a scratch pad normally. The tape head can read symbols from the tape one at a time and can move the tape to the left, right, or not move it, based on instructions from the control part.

The tape head tells the control the symbols it is scanning from the tape and the control decides what action to have the tape perform based on the scanned symbols and information (the program and data) stored in the control's memory.

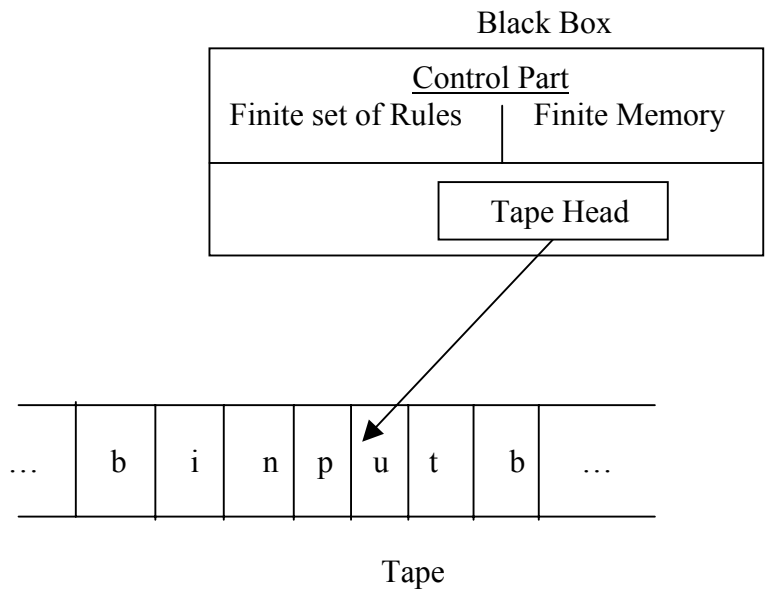

Figure. Schematic diagram of a Turing machine.

A set of input symbols is placed on the tape and the rules (program) in the control part are applied to produce an output set of symbols.

This process is analogous to elementary particle processes: an input set of particles interacts through various forces of nature and produces an output set of particles. The difference is that elementary particle processes are quantum probabilistic in nature. The laws of Physics (which appear to be finitely describable since they can be specified by the Standard Model lagrangian except for gravity) play the role of the finite set of rules.

**Quantum Probabilistic Grammars**
The major difference between Turing machine outputs and the outputs in particle physics are that output states in particle physics are quantum probabilistic. A given set of input symbols (particles) can produce a variety of output states with different probabilities calculable in the Standard Model. We need a Quantum Turing Machine to handle this more complex situation.

Quantum (and non-quantum) Turing Machines can be pictured in a convenient way by viewing the control part as containing a tape on which the rules are inscribed, the current state of the Turing machine is specified, and the current symbol being scanned by the tape head is stored.

The grammar rules of a Quantum Turing Machine are quantum probabilistic. In the simplest case each grammar rule has an associated number that we will call its relative probability amplitude. We will call this type of Quantum Probabilistic Grammar™ a *factorable Quantum Probabilistic Grammar*. The calculation of the probability for a transition from a specified input string to a specified output string is based on the following rules:

   1. The relative probability amplitude for an input string to be transformed to a specified output string is the sum of relative probability amplitudes for each possible sequence of transitions that leads from the input string to the output string.

   2. The relative probability amplitude for a sequence of grammar rule transitions is the product of the relative probability amplitudes of each transition.

   3. The relative probability for an input string to be transformed to an output string is the absolute value squared of the relative probability amplitude for the input string to be transformed to the output string.

   4. The absolute probability for an input string to be transformed to a specified output string is the relative probability for the input string to be transformed to the specified output string divided by the sum of the relative probabilities for the input string to be transformed to all possible output strings using the grammar rules. This rule guarantees the sum of the probabilities sums to one.

A Physics example is presented here in the book.

# 3

# The Standard Model Quantum Computer

**The Standard Model Quantum Turing Machine**
The Quantum Turing Machine that corresponds to the Standard Model has a number of exciting features that distinguish it from conventional Turing Machines.

First it accepts a language that has a finitely describable entangled Quantum Grammar™. Although the Standard Model has an entangled Quantum Grammar the grammar rules are finitely describable. Finitely describable means that the rules can be specified by a finite set of symbols. The rules generated from the interactions of the Standard Model are finite in number and each rule consists of a finite number of symbols. Thus the rules generated from the Standard Model are finitely describable.

A Quantum Turing Machine can be visualized as consisting of a control element and two tapes that play the role of computer memory. The control has a tape head that reads and writes symbols to the two tapes.

Tape I contains the specification of the grammar rules expressed as a finite string of symbols, the current state of the Quantum Turing Machine, and other data. Tape II contains the input string. After applying the grammar rules in a quantum probabilistic way an output state is generated. The output state is placed on Tape II in the simplest Quantum Turing Machine implementation.

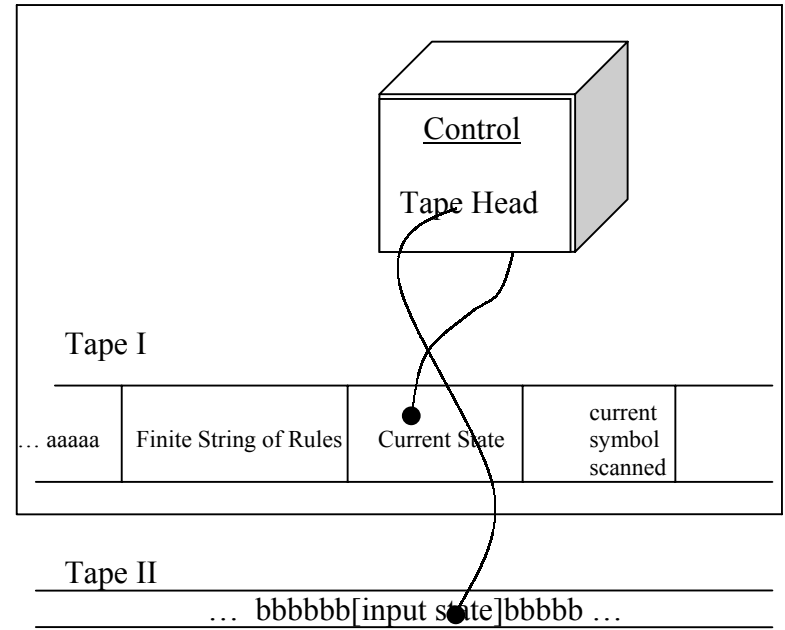

Figure. Quantum Turing Machine. Tape I plays the role of computer memory. Tape II is memory for input and output.

The behavior of a Quantum Turing Machine can be viewed as:

1. The Turing machine begins in the input state specified on tape II. An input string is placed on tape II. The other memory locations on tape II are filled with blank characters. In our case this string is a list of symbols for an input set of elementary particles that are about to interact. The Turing machine we are considering accepts any state consisting of a finite number of elementary particles. The connection between Turing machines and computer languages is brought out at this stage. A machine "accepts" a language if it can take any sentence (set of particles in our case) of the language, and perform a computation producing output (a set of output particles in our case). A Turing machine that accepts a language is an embodiment of the grammar of the language.

2. The Quantum Turing Machine applies the grammar rules to the input set of states in all possible ways to produce an output state that is a quantum superposition of states. Each possible output state has a certain probability of being produced.

3. The probability for producing a specified output state from a specified input state can be calculated as we illustrated in a simple example earlier using the relative probabilities associated with the Quantum Grammar™ rules of the modified $\phi^3$ theory.

4. The set of possible states of a Quantum Turing Machine[1] is infinite unlike non-Quantum Turing Machines that only have a finite number of possible states.

The Standard Model Quantum Turing Machine has some distinctive features:

1. Since the order of the particles in the input state string is not physically important we will consider the input string to be actually all permutations of the order of the particles in the input.

2. Since the Turing machine is quantum the rules are probabilistic in nature: a given set of input particles will in general produce many possible output particle states. Each output state will have a certain probability of being produced that can be calculated using the Standard Model.

3. The Quantum Grammar™ rules of the Standard Model Quantum Turing Machine have internal symmetries that result in symmetries in the input and output states.

4. Since the momenta and spins of the input and output particles are physically very important the Standard Model Quantum Turing Machine must take account of these properties in the input and output states as well as internally when calculating transition probabilities.

So we must picture the input particle state on tape II as containing not only the particle symbol but also momenta and spin data.

To get an idea of how a Quantum Turing Machine would take an input set of particles and produce a set of output particles we will consider the case of two electrons colliding with such energy that an electron-positron pair is created:

$$ee \rightarrow eepe$$

where e represents an electron and p represents a positron (the electron's antiparticle). One of the corresponding Feynman-like diagrams is:

---

[1] D. Deutsch, Proceedings of the Royal Society of London, A **400** 97 (1985) describes (universal) Quantum Computers and points out they can simulate continuous physical systems because they have a continuum (infinite number) of possible states. As page 107 points out "a quantum computer has an infinite-dimensional state space". Quantum Computers are equivalent to Quantum Turing Machines as we will see.

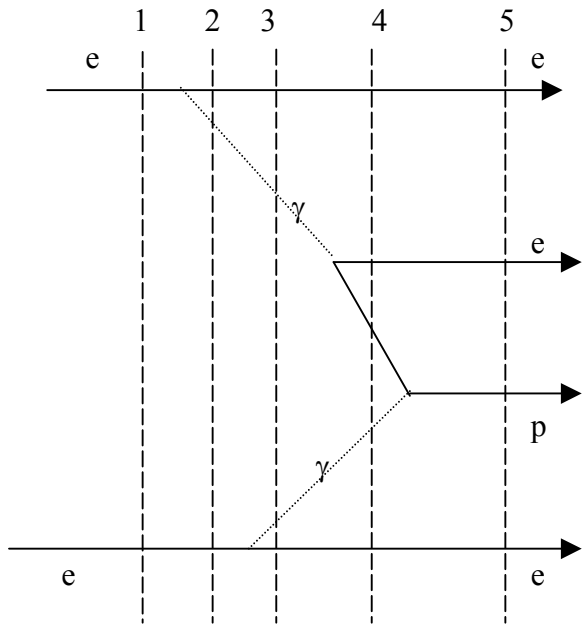

where γ represents a photon. The input string can change according to the grammar rules in the following way:

$$ee \rightarrow e\gamma e \rightarrow e\gamma\gamma e \rightarrow eep\gamma e \rightarrow eepe$$

Since the Quantum Turing Machine is probabilistic there are many – in fact an infinite number – of ways in which the transition

$$ee \rightarrow eepe$$

can take place – each with its own probability of happening. The sample sequence of transitions shown above is only one of these possible ways. The total probability of this transition is the square of the sum of the probability amplitudes for all possible ways according to quantum mechanics. Nature requires us to take account of all possible ways of transitioning from the input state of particles to the output state of particles. The total probability of the output state being produced is a sum of the contributions of all the possible alternate ways of reaching that output state.

In addition, electron-electron scattering can produce many other output states depending on the initial energy of the electrons. Each output state has its own probability of occurring. Some examples are:

$$ee \rightarrow eq\underline{q}e$$

$$ee \rightarrow eepepe$$

$$ee \rightarrow e\mu\underline{\mu}e$$

$$ee \rightarrow e\mu\underline{\mu}epe$$

where q represents a quark, $\underline{\mu}$ represents the muon antiparticle and $\underline{q}$ represents an antiquark.

The Quantum Turing Machine representation does raise several interesting prospects for the theory of elementary particles embodied in the Standard Model. First, the Quantum Turing Machine representation raises the possibility that some of the powerful techniques and general results of the theory of computation can be brought over to physics and perhaps provide guidance on the next stage after the Standard Model.

Secondly, and perhaps more importantly, the separation of the input and output states (they are on tape II) from the intermediate calculational states of the Turing machine (that are on tape I) is suggestive of a somewhat different approach to the fundamentals of particle interactions: The space-time of the incoming and outgoing particles may be different from the "space-time" describing the interactions and internal structure of the interacting particles. This view is based on thinking of tapes I and II as representing separate space-times.

A precursor of this point of view appears in Quantum Field Theory. In Quantum Field Theory the interaction of particles is viewed as consisting of three phases: an initial state where the particles are widely separated and distinct, an interaction region where the particles "collide" and interact perhaps creating new particles, and a final state where the outgoing particles are widely separated.

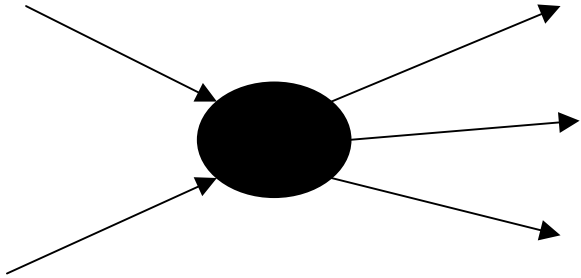

Figure. Two particles collide and generate a three particle outgoing state in Quantum Field Theory.

In conventional Quantum Field Theory the space-time in the interacting region is conventionally assumed to be the same as the space-time of the incoming and outgoing states. Nevertheless Quantum Field Theory distinguishes the interaction region from the region of the incoming and outgoing particles.

The SuperString approach to the theory of elementary particles introduces a separate space-time to describe the elementary particles. Elementary particles are viewed as strings vibrating in this space-time. Can one view the SuperString space-time as tape I and the external behavior of the elementary particles taking place in normal space-time as tape II? Perhaps the Quantum Turing Machine representation of the Standard Model is the key to the next level of our understanding of elementary particles and Nature.

# 4

# Quantum Computer Processor Operations and Quantum Computer Languages

## Introduction

A natural question that arises when one considers Quantum Computers is the role of the Quantum Computer processor and the operations it supports. A further question of some interest is whether a quantum machine language exists and what its nature might be. Lastly the question of higher level languages is also relevant. Can we develop a Quantum Assembly Language™? What is the nature of High Level Quantum Languages™? Are there, for example, equivalents to the C or C++ languages?

## Computer Machine and Assembly Languages

The traditional (non-quantum) computer can be viewed simply as a main memory, an accumulator or register (modern computers have many registers), and a central processing unit (CPU) that executes a program (instructions) step by step. It can be visualized as:

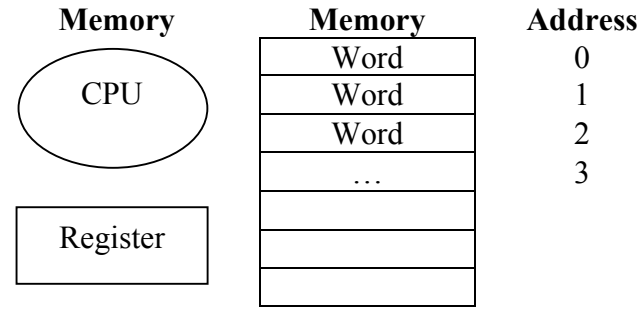

Figure. Simplified model of a normal computer.

A set of data and a program (or set of instructions) is stored in memory and the CPU executes the program step by step using the data to produce an output set of data.

The basic instructions of assembly language and machine language move data values between memory and the register (or registers), manipulate the data value in the register and provide basic arithmetic and logical operations[2]:

    LOAD M   –   load the value at memory location M into the register

    STORE M   – store the value in the register at memory location M

    SHIFT k   –   shift the value in the register by k bits

---

[2] See for example Kurt Maly and Allen R. Hanson, *Fundamentals of the Computing Sciences* (Prentice-Hall, Inc., Englewood Cliffs, NJ, 1978) Chap. 8.

The following arithmetic instructions modify the value in the register. The AND, OR and NOT instructions perform bit-wise and, or and not operations.

    ADD M    – add the value at memory location M to the value in the register

    SUBTRACT M   – subtract the value at memory location M from the value in the register

    MULTIPLY M   – multiply the value in the register by the value at memory location M

    DIVIDE M   – divide the value in the register by the value at memory location M

    AND M –    change the value in the register by "anding" it with the value at memory location M

    OR M –    change the value in the register by "oring" it with the value at memory location M

    NOT –    change the value in the register by "not-ing" it

The following instructions implement input and output of data values.

    INPUT M  – input a value storing it at memory location M

    OUTPUT M – output the value at memory location M

A computer has another register called the Program Counter. The value in the program counter is the memory location of the next instruction to execute. The following instructions support non-sequential flow of control in a program. A program can "leap" from one instruction in a program to another instruction many steps away and resume normal sequential execution of instructions.

    TRA M   – set the value of the program counter to the value at memory location M

    TZR M   – set the value of the program counter to the value at memory location M if the value in the register is zero.

    HALT    – stop execution of the program

The above set of instructions form an extremely simple assembly language. They also are in a one-to-one correspondence with machine instructions (machine language). Most assembly and machine languages have a much more extensive set of instructions.

### Algebraic Representation of Assembly Languages

The normal view of assembly language is that it has a word or instruction oriented format. Some assembly language programmers would even say that assembly language is somewhat English-like in part.

Computer languages in general have tended to become more English-like in recent years in an attempt to make them easier for programmers. Some view a form of highly structured English to be a goal for computer programming languages.

In this section we follow the opposite course and show that computer languages can be reduced to an algebraic representation. By algebraic we mean that the computer language can be represented with operator expressions using operators that have an algebra similar to that of the raising and lowering operators seen earlier. We will develop the algebraic representation for the case of the simple assembly language described in the previous section. There are a number of reasons why this reduction is interesting:

    1. It may help to understand SuperString dynamics more deeply (later in this chapter).

2. It will deepen our understanding of computer languages.

3. It provides a basis for the understanding of Quantum Computers.

4. It may have a role in research on one of the major questions of computer science: proving a program actually does what it is designed to do. Algebraic formalisms are generally easier to prove theorems then English-like formalisms.

The algebraic representation can be defined at the level of individual bits based on anti-commuting Fermi operators. But it seems more appropriate to develop a representation for "words" consisting of some number of bits. An algebraic representation for a word-based assembly language can be developed using commuting harmonic oscillator-like raising and lowering operators.

A word consists of a number of bits. In currently popular computers the word size is 32 bits (32-bit computer). The size of the word determines the largest and smallest integer that can be stored in the word. The largest integer that can be stored in a 32-bit word is 4,294,967,294 and the smallest integer that can be stored in a 32-bit word is 0 if we treat words as holding unsigned integers.

To develop a simple algebraic representation of assembly language we will assume the size of a word is so large that it can be viewed as infinite to a good approximation. (It is also possible to develop algebraic representations for finite word sizes.) As a result memory locations can contain non-negative integers of arbitrarily large value.

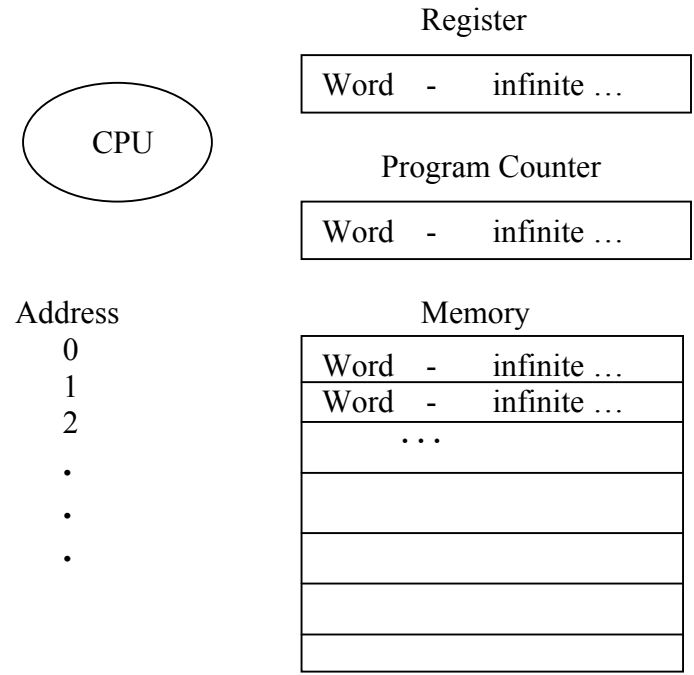

Figure. Visualization of a Computer with infinite words.

To establish the algebraic representation we associate a harmonic raising operator $a_i^\dagger$ and a lowering operator $a_i$ with each memory location. These operators satisfy the commutation relations:

$$[\, a_i, a_j^\dagger \,] = \delta_{ij}$$
$$[\, a_i, a_j \,] = 0$$
$$[\, a_i^\dagger, a_j^\dagger \,] = 0$$

where $\delta_{ij}$ is 1 if i = j and zero otherwise. We define a pair of raising and lowering operators for the register r and $r^\dagger$ with commutation relations

$$[\, r_i, r_j^\dagger \,] = \delta_{ij}$$
$$[\, r_i, r_j \,] = 0$$
$$[\, r_i^\dagger, r_j^\dagger \,] = 0$$

The ground state of the computer is the state with the values at all memory locations set to zero. It is represented by the vector

$$| 0, 0, 0, \ldots \rangle \equiv | 0 \rangle \equiv \Phi_V$$

A state of the computer will be represented by a vector of the form

$$| n, m, p, \ldots \rangle = N (r^\dagger)^n (a_0^\dagger)^m (a_1^\dagger)^p \ldots | 0 \rangle$$

where N is a normalization constant and with the first number being the value in the register, the second number the value at memory location 0, the third number the value at memory location 1, and so on. For simplicity we will not consider superpositions of computer states at this point. We will discuss superpositions later in this chapter. Within this limitation we can set a computer state to have certain initial values in memory and then have it evolve by executing a "program" to a final computer state with a different set of computer values in memory. The "program" is a mapping of the instructions of an assembly language program to algebraic expressions in the raising and lowering operators.

## Basic Operators of the Algebraic Representation

The key operators that are required for the algebraic representation are:

Fetch the Value at a Memory Location (Number Operator)

$$N_m = a_m^\dagger a_m$$

For example,

$$N_m | \ldots, n, \ldots \rangle = n | \ldots, n, \ldots \rangle$$

↑ $m^{th}$ memory location value

Set the Value at Memory Location m to Zero

$$M_m = \frac{(a_m)^{N_m}}{\sqrt{N_m!}}$$

The above expression for $M_m$ is symbolic. The expression represents the following expression in which the operators are carefully ordered to avoid complications (c-numbers etc.) resulting from reordering.

$$M_m \equiv \sum_q \frac{(\ln a_m)^q N_m^q}{q!} \frac{1}{\sqrt{N_m!}}$$

where the sum ranges from 0 to ∞. When $M_m$ is applied to a state it sets the value of the $m^{th}$ memory location to zero.

$$M_m | \ldots, n, \ldots \rangle = \frac{(a_m)^n}{\sqrt{n!}} | \ldots, n, \ldots \rangle$$

↑ $m^{th}$ memory location value

$$= | \ldots, 0, \ldots \rangle$$

The repeated application of factors of $N_m$ to the state results in factors of n.

Change the Value at Memory Location m from 0 to the Value at Location n

$$P_m^n = \frac{(a_m^\dagger)^{N_n}}{\sqrt{N_n!}}$$

The above expression for $P_m^n$ is also symbolic. The expression represents the following expression in which the operators are carefully ordered to avoid complications (c-numbers etc.) resulting from reordering.

$$P_m^n \equiv \sum_q \frac{(\ln a_m^\dagger)^q N_n^q}{q!} \frac{1}{\sqrt{N_n!}}$$

where the sum ranges from 0 to ∞. When $P_m^n$ is applied to a state it changes the value of the $m^{th}$ memory location from zero to the value at the $n^{th}$ memory location.

$$P_m^n | \ldots, \underset{m^{th}}{0}, \ldots, \underset{n^{th}}{x}, \ldots > = \frac{(a_m^\dagger)^x}{\sqrt{x!}} | \ldots, 0, \ldots, x, \ldots >$$

$$= | \ldots, x, \ldots, x, \ldots >$$

The application of the factors of $N_n$ to the state results in factors of x that lead to the above expression when summed.

The operators $M_m$ and $P_m^n$ enable us to simply express the algebraic equivalent of assembly language instructions:

**LOAD m** – load the value at memory location m into the register

$$P_r^m M_r$$

**STORE m** – store the value in the register at memory location m

$$P_m^r M_m$$

**SHIFT k** – shift the value in the register by k bits. If k is positive the bit shift is to the right and if k is negative the bit shift is to the left. The bits are numbered from the leftmost bit which is bit 0 corresponding to $2^0$. The next bit is bit 1 corresponding to $2^1$ and so on.

If the bit shift is to the right (k > 0) then we assume the padding bits are 0's. For example a shift of the bit pattern for 7 = 1110000 … one bit to the right is 14 = 01110000 … As a result the value in the register is doubled (k = 1), quadrupled (k = 2), and so on. The algebraic expression for a k bit right shift is

$$\sum_q \frac{(\ln a_r^\dagger)^q S_r^q}{q!} \frac{\sqrt{N_r!}}{\sqrt{T_r!}}$$

where

$$S_r = (2^k - 1)N_r$$

and

$$T_r = 2^k N_r$$

If the bit shift is to the left (negative k), then we assume zero bits are added "at ∞". If k = -1 then the effect of left shift is to divide the value in the register by two (dropping the fractional part). If k = -2 then the effect of left shift is to divide the value in the register by four (dropping the fractional part) and so on. The algebraic expression that implements left shift is

$$\sum_q \frac{(\ln a_r)^q U_r^q}{q!} \frac{\sqrt{N_r!}}{\sqrt{V_r!}}$$

where

$$U_r = N_r - [2^k N_r]$$

and

$$V_r = [2^k N_r]$$

with [ z ] being the value of z truncated to an integer (fractional part dropped).

**ADD m** – add the value at memory location m to the value in the register

$$\sum_q \frac{(\ln a_r^\dagger)^q N_m^q}{q!} \frac{\sqrt{N_r!}}{\sqrt{(N_r + N_m)!}}$$

**SUBTRACT m** – subtract the value at memory location m from the value in the register (assumes the value in the register is greater than or equal to the value at location m)

$$\sum_q \frac{(\ln a_r)^q N_m^q}{q!} \frac{\sqrt{N_r!}}{\sqrt{(N_r - N_m)!}}$$

**MULTIPLY m** – multiply the value in the register by the value at memory location m

$$\sum_q \frac{(\ln a_r^\dagger)^q (N_r)^q (N_m - 1)^q}{q!} \frac{\sqrt{N_r!}}{\sqrt{(N_r N_m)!}}$$

**DIVIDE m** – divide the value in the register by the value at memory location m

$$\sum_q \frac{(\ln a_r^\dagger)^q W^q}{q!} \frac{\sqrt{N_r!}}{\sqrt{X!}}$$

where

$$W = N_r - [ N_r / N_m ]$$

and

$$X = [ N_r / N_m ]$$

with [ z ] being the value of z truncated to an integer (fractional part dropped).

**AND m** – change the value in the register by "and-ing" it with the value at memory location m

$$\sum_q \frac{((\ln a_r^\dagger)^q (W)^q \theta (W) + (\ln a_r)^q (-W)^q \theta (-W))}{q!} \frac{\sqrt{N_r!}}{\sqrt{X!}}$$

where

$$W = N_r \& N_m - N_r$$

and where

$$X = N_r \& N_m$$

with $\theta(z) = 1$ if $z > 0$ and 0 if $z < 0$. The & operator (adopted from the C programming language) performs bitwise AND. Corresponding bits in each operand are "multiplied" together using the multiplication rules:

$$1 \& 1 = 1$$

$$1 \& 0 = 0 \& 1 = 0 \& 0 = 0$$

For example the binary numbers 1010 & 1100 = 1000 or in base 10 5 & 3 = 1.

**OR m** – change the value in the register by "or-ing" it with the value at memory location m

$$\sum_q \frac{(\ln a_r^\dagger)^q (W)^q}{q!} \frac{\sqrt{N_r!}}{\sqrt{X!}}$$

where

$$W = N_r \mid N_m - N_r$$

and where

$$X = N_r \mid N_m$$

The | operator (adopted from the C programming language) performs bitwise OR. Corresponding bits in each operand are "multiplied" together using the multiplication rules:

$$1 \mid 1 = 1 \mid 0 = 0 \mid 1 = 1$$

$$0 \mid 0 = 0$$

For example the binary numbers 1010 | 1100 = 1110 or in base 10 5 | 3 = 7.

**NOT** – change the value in the register by "noting" it

$$\sum_q \frac{((\ln a_r^\dagger)^q (W)^q \theta(W) + (\ln a_r)^q (-W)^q \theta(-W))}{q!} \frac{\sqrt{N_r!}}{\sqrt{X!}}$$

where

$$W = \sim N_r - N_r$$

and where

$$X = \sim N_r$$

with $\theta(z) = 1$ if $z > 0$ and $0$ if $z < 0$. The ~ operator (adopted from the C programming language) performs bitwise NOT. Each 1 bit is replaced by a 0 bit and each 0 bit is replaced by a 1 bit. Since we have infinite words in our computer we supplement this rule with the restriction that the exchange of 1's and 0's only is made up to and including the rightmost 1 bit in the operand. The 0 bits beyond that remain 0 bits. For example the binary number ~101 = 010 or in base 10, ~3 = 2.

**INPUT m** – input a value storing it at memory location m. The input device is usually associated with a memory location from which the input symbolically takes place. We will designate the memory location of the input device as *in*.

$$P_m{}^{in} M_m$$

**OUTPUT m** – output the value at memory location m. The output device is usually associated with a memory location to which output symbolically takes place. We will designate the memory location of the output device as *out*.

$$P_{out}{}^m M_{out}$$

**TRA m** – set the value of the program counter to the value at memory location m. If we designate the program counter memory location as pc then this instruction is mapped to

$$P_{pc}{}^m M_{pc}$$

**TZR m** – set the value of the program counter to the value at memory location m if the value in the register is zero.

$$(P_{pc}{}^m M_{pc})^{\theta(N_r)\,\theta(-N_r)}$$

using $\theta(0) = 1$.

**HALT** – stop execution of the program. The halt in a program is mapped to a "bra" state vector.

$$< \ldots |$$

## A Simple Assembly Language Program

Assembly language instructions can be combined to form an assembly language program. Perhaps the best way to see how the algebraic representation of assembly language works is to translate a simple assembly language program into its algebraic equivalent.

The program that we will consider is:

1    INPUT x
2    INPUT y
3    LOAD x
4    ADD y
5    STORE z
6    OUTPUT z
7    HALT

This program translates to the algebraic equivalent:

$$\begin{array}{ccccccc} 7 & 6 & 5 & 4 & 3 & 2 & 1 \quad \text{Steps} \\ < \ldots | P_{out}^z M_{out} & P_z^r M_z & \frac{(a_r^\dagger)^{N_y} \sqrt{N_r!}}{\sqrt{(N_r + N_y)!}} & P_r^x M_r & P_y^{in} M_y & P_x^{in} M_x | \ldots > \end{array}$$

where the power of $a_r^\dagger$ is represented by a power series expansion as seen earlier.

The algebraic expression in the brackets produces one output state from a given initial state. The values in memory after the last step correspond to one and only one output state of the form:

$$< n, m, p, \ldots | = (N (r^\dagger)^n (a_0^\dagger)^m (a_1^\dagger)^p \ldots | 0 >)^\dagger$$

where N is a normalization constant.

This simple program does not produce a superposition of states. As a result programs of this type are analogous to ordinary programs for normal, non-Quantum computers. The numbers in memory after the program concludes are the "output" of the program. We will see programs in succeeding sections that take a computer of fixed state $N (r^\dagger)^n (a_0^\dagger)^m (a_1^\dagger)^p \ldots | 0 >$ and produce a superposition of states that must be interpreted quantum mechanically. These programs are quantum in nature and the computer that runs them must be a quantum computer.

## Programs and Program Logic

The simple program of the last section corresponded to a sequential program that executed step by step. We now turn to more complex programs with program logic that supports non-sequential execution of programs. When this type of program executes the execution of the instructions can lead to jumps from one instruction to another instruction in another part of the program.

Programs are linear – one instruction executes after another. But they are not sequential – the instructions do not always execute step by step sequentially. A program can specify jumps ("goto" instructions) in the code from the current instruction to an instruction several steps after the current instruction or several steps back to a previous instruction. The code then executes sequentially until the next jump is encountered.

These jumps in the code at the level of assembly language implement the control constructs such as goto statements, if expressions, for loops, and switch expressions seen in higher level languages such as C and C++.

Jumps in code can be implemented in the algebraic representation of programs by having a program counter memory value that increments as the algebraic factor corresponding to each step executes. Steps in the program can execute or not execute depending on the current value of the program counter.

Changes in the program counter value are made using the TRA and TZR instructions. In the algebraic representation the program counter variable can be used to manage the execution of the program steps.

The key algebraic constructs supporting non-sequential program execution are:

Execute instruction only if PC ≤ n

$$( \ldots )^{\theta(n - N_{pc})}$$

Execute instruction only if PC ≥ n

$$( \ldots )^{\theta(N_{pc} - n)}$$

Execute instruction only if PC = n

$$( \ldots )^{\theta\theta(N_{pc} - n)}$$

Execute instruction only if PC not equal to n

$$( \ldots )^{\theta(N_{pc} - n) + \theta(n - N_{pc}) - 2\theta\theta(N_{pc} - n)}$$

where the parentheses contain one or more instructions and where $\theta\theta(x) = 1$ if $x = 0$ and zero otherwise. The function $\theta\theta(x)$ can be represented by step functions as

$$\theta\theta(x) = \theta(x)\,\theta(-x)$$

Using these constructs we can construct non-sequential programs that supprt "goto's", if's and other control constructs seen in higher level languages.

To illustrate this feature of the algebraic representation we will consider an enhancement of the assembly language program seen earlier:

1    INPUT x
2    INPUT y
3    LOAD x
4    TZR y
5    ADD y
6    STORE z
7    OUTPUT z
8    HALT

This program has the new feature that if the first input – to memory location x – is zero, then instruction 4 will cause a jump to the instruction specified by the value stored at memory location y.

For example if the inputs are 0 placed at memory location x and 2 placed at memory location y, then the TZR instruction will cause the program to jump to instruction 2 from instruction 4. Then the program will proceed to execute from instruction 2.

Another example of a case with a jump is if the input to memory location x is zero and the input to memory location y is 6 then the program jumps from instruction 4 to instruction 6 and the program completes execution from there. If the input to memory location x is non-zero no jump takes place.

To establish the algebraic equivalent of the preceding example we have to use the non-sequential constructs provided earlier in this section. In addition we must define the equivalent recursively because of the possibility that the program may jump backwards to an earlier instruction in the program. If only "forward" jumps were allowed then recursion would not be needed.

An algebraic representation of the program that supports only forward leaps is:

$$\overset{8}{<\ldots|}\;\overset{7}{(a_{pc}^\dagger P_{out}^z M_{out})^{\theta\theta(N_{pc}-7)}}$$

$$\overset{6}{(a_{pc}^\dagger P_z^r M_z)^{\theta\theta(N_{pc}-6)}}$$

$$\overset{5}{(\frac{a_{pc}^\dagger (a_r^\dagger)^{N_y} \sqrt{N_r!}}{\sqrt{(N_r+N_y)!}})^{\theta\theta(N_{pc}-5)}}$$

$$\overset{4}{(a_{pc}^\dagger)^{1-\theta\theta(N_r)} (P_{pc}^y M_{pc})^{\theta\theta(N_r)\,\theta\theta(N_{pc}-4)}}$$

$$\overset{3}{a_{pc}^\dagger P_r^x M_r}$$

$$\overset{2}{a_{pc}^\dagger P_y^{in} M_y}$$

$$\overset{1}{a_{pc}^\dagger P_x^{in} M_x}$$

$$a_{pc}^\dagger M_{pc}\;|\ldots>$$

The program steps are numbered above each corresponding expression. The step function expressions enable the jump to take place successfully.

A program with forward and backward jumps supported requires a recursive definition. We will define the recursive function f() with:

$$f() = \overset{7}{(a_{pc}^\dagger P_{out}^z M_{out})^{\theta\theta(N_{pc}-7)}}$$

$$\overset{6}{(a_{pc}^\dagger P_z^r M_z)^{\theta\theta(N_{pc}-6)}}$$

$$\overset{5}{(\frac{a_{pc}^\dagger (a_r^\dagger)^{N_y} \sqrt{N_r!}}{\sqrt{(N_r+N_y)!}})^{\theta\theta(N_{pc}-5)}}$$

$$\overset{4}{(a_{pc}^\dagger)^{1-\theta\theta(N_r)} (f() P_{pc}^y M_{pc})^{\theta\theta(N_r)\,\theta\theta(N_{pc}-4)}}$$

$$\overset{3}{(a_{pc}^\dagger P_r^x M_r)^{\theta\theta(N_{pc}-3)}}$$

$$\overset{2}{(a_{pc}^\dagger P_y^{in} M_y)^{\theta\theta(N_{pc}-2)}}$$

$$\overset{1}{(a_{pc}^\dagger P_x^{in} M_x)^{\theta\theta(N_{pc}-1)}}$$

The program is

$$f() a_{pc}^\dagger M_{pc}\;|\ldots>$$

This program is well behaved except if the input value placed at the y memory location is 4. In this case the program recursively executes forever. This defect can be removed by using another memory location for a counter variable.

We can modify the program so that the program only recursively calls itself a finite number of times by having each recursive call decrease the counter variable by one. When the value reaches zero the recursion terminates. An example of such a program (set to allow at most 10 iterations of the recursion) is:

$$g() = (a_{pc}^\dagger P_{out}^z M_{out})^{\theta\theta(N_{pc}-7)} \quad 7$$

$$(a_{pc}^\dagger P_z^r M_z)^{\theta\theta(N_{pc}-6)} \quad 6$$

$$(\frac{a_{pc}^\dagger (a_r^\dagger)^{N_y} \sqrt{N_r!}}{\sqrt{(N_r + N_y)!}})^{\theta\theta(N_{pc}-5)} \quad 5$$

$$(a_{pc}^\dagger)^{1-\theta\theta(N_r)} ((a_{pc}^\dagger)^{1-\theta(N_w)}(g())^{\theta(N_w)} a_w P_{pc}^y M_{pc})^{\theta\theta(N_r)\,\theta\theta(N_{pc}-4)} \quad 4$$

$$(a_{pc}^\dagger P_r^x M_r)^{\theta\theta(N_{pc}-3)} \quad 3$$

$$(a_{pc}^\dagger P_y^{in} M_y)^{\theta\theta(N_{pc}-2)} \quad 2$$

$$(a_{pc}^\dagger P_x^{in} M_x)^{\theta\theta(N_{pc}-1)} \quad 1$$

The program is

$$g() a_{pc}^\dagger M_{pc} (a_w^\dagger)^{10} M_w \,|\, \ldots >$$

where w is some memory location. We conjecture that any assembly language program using the previously specified instructions can be mapped to an algebraic representation – possibly with the use of additional memory for variables such as the counter variable seen above.

Using the algebraic constructs supporting non-sequential program execution we can create algebraic representations of assembly language programs. These programs have a definite input state and through the execution of the program they evolve into a definite output state -–not a superposition of output states. Therefore they faithfully represent assembly language programs. On the other hand they are quantum in the sense that they use states and harmonic oscillator-like raising and lowering operators. The types of programs we are creating in this approach are "sharp" on the space of states. One input state evolves through the program's execution to one and only one output state with probability one.

These types of programs are analogous to free field theory in which incoming particles evolve without interaction to an output state containing the same particles.

In the next section we extend the ideas in this section to quantum programming where a variety of output states are possible – each with a certain probability of being produced.

**Quantum Assembly Language™ Programs**

In this section we will first look at a simplified quantum program that illustrates quantum effects but in actuality is a sum of deterministic assembly language programs mapped to algebraic equivalents. Consider a "quantum" program that is the sum of three ordinary programs $g_1()$, $g_2()$ and $g_3()$ of the type seen in the last section. Further let us assume the set of orthonormal states

$$|\,n, m, p, \ldots >$$

that we saw in the previous sections with

$$< X\,|\,Y > = \delta_{XY}$$

where $\delta_{XY}$ represents a product of Kronecker $\delta$ functions in the individual values in memory of the $|\,X >$ and $|\,Y >$ states. Further let us assume

$$| n_1, m_1, p_1, \ldots > = g_1()| \ldots >$$

$$| n_2, m_2, p_2, \ldots > = g_2()| \ldots >$$

$$| n_3, m_3, p_3, \ldots > = g_3()| \ldots >$$

for some initial state of the quantum computer. Then

$$\alpha g_1() + \beta g_2() + \gamma g_3()| \ldots >$$

is a "quantum" program where $\alpha$, $\beta$, and $\gamma$ are constants such that

$$|\alpha|^2 + |\beta|^2 + |\gamma|^2 = 1$$

The quantum program produces the state $| n_1, m_1, p_1, \ldots >$ with probability $|\alpha|^2$, the state $| n_2, m_2, p_2, \ldots >$ with probability $|\beta|^2$, and the state $| n_3, m_3, p_3, \ldots >$ with probability $|\gamma|^2$.

We now have a quantum probabilistic computer. The programs $g_1()$, $g_2()$ and $g_3()$ are being executed in *parallel* in a quantum probabilistic manner.

Currently, the most feasible way of creating a Quantum Computer with current technology or reasonable extrapolations of current technology is to create a material which approximates a lattice with spins at each lattice site that we can orient electromagnetically at the beginning of a program. The execution of a program takes place by applying electromagnetic fields that have a time dependence specific to the computation. The electromagnetic fields implement a custom-tailored set of interactions between the spins in the material that simulates the calculation to be performed.

The interactions are specified with some Hamiltonian or some effective Hamiltonian and the initial state of the lattice spins evolves dynamically to some configuration that is then measured.

The Hamiltonians are normally specified using the space-time formalism that is a familiar part of Quantum Mechanics. A Hamiltonian specifies the time evolution of a system starting from an initial state. We can introduce an explicit time dependence in states by using the notation:

$$| \Psi(t) >$$

to denote the state of a Quantum Computer at time t. The general state of the computer at time t can be written as a superposition of the number representation states:

$$| \Psi(t) > = \sum_n f_n(t)| n_1, n_2, n_3, \ldots >$$

where n represents a set of values $n_1, n_2, n_3, \ldots$

The time evolution of the states can be specified using the Hamiltonian operator H as

$$| \Psi(t) > = e^{-iHt}| \Psi(0) >$$

With this Hamiltonian formulation we can imagine wishing to simulate a physical (or mathematical) process, defining a Hamiltonian that corresponds to the process, and then creating an experimental setup using a set of lattice spins in some material that implements the simulation. The experimental setup will prepare the initial state of the spins, create a fine tuned interaction that simulates the physics of the process, and then, after the system has evolved, will measure the state of the system at time t. Repeated performance of this procedure will determine the probability distribution associated with the final state of the Quantum Computer. The probability distribution is specified by $|f_n(t)|^2$ as a function of the sets of numbers denoted by n.

A simple example of a Hamiltonian that causes a Quantum Computer to evolve in a non-trivial way is:

$$H = \sum_{m=0}^{\infty} a_{m+1}^\dagger a_m$$

(This example was chosen partly because it has a form similar to a Virasoro algebra generator in SuperString Theory.) Let us assume the initial state of the Quantum Computer at t = 0 is

$$| 1, 0, 0, 0, \ldots >$$

that is, an initial value of 1 in the first word in memory and zeroes in all other memory locations. At time t the state of memory is:

$$| \Psi(t) > = \sum_{n=0}^{\infty} f_n(t) | 0, 0, \ldots, \underset{n^{th}\ memory\ location}{1}, 0, \ldots >$$

with

$$f_n(t) = (-it)^n/n!$$

using the power series expansion of the exponentiated Hamiltonian expression. The probability of finding the state

$$| 0, 0, \ldots, \underset{n^{th}\ memory\ location}{1}, 0, \ldots >$$

is

$$(t^n/n!)^2$$

At first glance the Hamiltonian approach is very different from the Quantum Assembly Language™ approach discussed above. However these approaches can be interrelated in special cases and (we conjecture) in the general case through sufficiently clever transformations. For example, the preceding Hamiltonian can be re-expressed as assembly language instructions

$$H = \sum_{m=0}^{\infty} (STORE\ (m+1))(ADD\ "1")(LOAD\ (m+1)) \cdot$$

$$\cdot (STORE\ m)(SUBTRACT\ "1")(LOAD\ m)$$

where a value is loaded into the register from memory location m and then 1 is added to the value in the register. The "1" expression represents a literal value one not a memory location. The parentheses around m+1 indicates it is the $(m+1)^{th}$ memory location – not the addition of one to the value at the $m^{th}$ location.

The preceding assembly language expression for H can be replaced with the algebraic representation expression:

$$H = \sum_{m=0}^{\infty} P_{m+1}^{\ r} M_{m+1} a_r^\dagger \frac{1}{\sqrt{(N_r + 1)}} P_r^{\ m+1} M_r P_m^{\ r} M_m a_r \sqrt{N_r} P_r^{\ m} M_r$$

This complex expression is not an improvement in one sense. The original Hamiltonian expression was much simpler. Its importance is the mapping that it embodies from a quantum mechanical Hamiltonian to an assembly language expression to an algebraic representation of the assembly language.

If we regard the value in the register as a "scratchpad" value as programmers often do, then we can establish a representation of $a_m^\dagger$ and $a_m$ in terms of the algebraic representation of assembly language instructions.

$$a_m^\dagger \equiv P_m^{\ r} M_m a_r^\dagger \frac{1}{\sqrt{(N_r + 1)}} P_r^{\ m} M_r$$

and

$$a_m \equiv P_m^{\ r} M_m a_r \sqrt{N_r} P_r^{\ m} M_r$$

The power series expansion of the exponentiated Hamiltonian in the previous example is an example of the use of Perturbation Theory. The direct solution of a problem is often not feasible because of the complexity of the dynamics. Physicists have a very well developed theory for the approximate solution of these difficult problems called Perturbation Theory. Perturbation Theory takes an exact solution of a simplified version of the problem and then calculates corrections to that solution that approximate the exact solution of the problem. In the preceding example the initial state of the Quantum Computer represents a time-independent description of the Quantum Computer. The time-dependent description of the Quantum Computer which is the sought-for solution requires the evaluation of the result of the application of the exponentiated Hamiltonian to the initial state. For a small elapsed time, the exponential can be expanded in a power series and the application of the first few terms of the power series to the initial state approximates the actual state of the Quantum Computer. Thus we have a Perturbation Theory for the time evolution of the Quantum Computer expressed as an expansion in powers of the elapsed time.

## Bit-Level Quantum Computer Language

In the previous section we examined a Quantum Assembly Language™ with words consisting of an infinite sets of bits. In this section we will examine the opposite extreme – a Quantum Computer Language with one-bit words. One can also create Quantum Computer Languages for intermediate cases such as 32-bit words.

A Bit-Level Quantum Computer Language can be represented with anti-commuting Fermi operators $b_i$ and $b_i^\dagger$ for $i = 0, 1, 2, \ldots$ representing each bit location in the Quantum Computer's memory with the anti-commutation rules:

$$\{b_i, b_j^\dagger\} = \delta_{ij}$$

$$\{b_i, b_j\} = 0$$

$$\{b_i^\dagger, b_j^\dagger\} = 0$$

where $\delta_{ij}$ is 1 if $i = j$ and zero otherwise. We will assume an (unrealistic) one-bit register with a pair of raising and lowering operators $r$ and $r^\dagger$ for the register with the anti-commutation relations:

$$\{r_i, r_j^\dagger\} = \delta_{ij}$$

$$\{r_i, r_j\} = 0$$

$$\{r_i^\dagger, r_j^\dagger\} = 0$$

The ground state of the computer is the state with the values at all bit memory locations set to zero. It is represented by the vector

$$|0, 0, 0, \ldots> \equiv |0> \equiv \Phi_V$$

A typical state of the computer will be represented with a vector such as

$$|1, 1, 1, \ldots> = r^\dagger b_0^\dagger b_1^\dagger \ldots |0>$$

with the first number being the value in the register, the second number the value at memory location 0, the third number the value at memory location 1, and so on.

A specified Quantum Computer state evolves as a Quantum Computer Program executes to a final computer state. A Bit-Level Quantum Computer Program can be represented as an algebraic expression in anti-commuting raising and lowering operators. The approach is similar to the approach seen earlier in this chapter for infinite-bit words using commuting operators.

**Basic Operators of the Bit-Level Quantum Language**

The key operators that are required for the algebraic representation of a Bit-Level Quantum Computer Language™ are:

Fetch the Value at a Memory Location (Number Operator)

$$N_m = b_m^\dagger b_m$$

For example,

$$N_m | \ldots, 1, \ldots \rangle = | \ldots, 1, \ldots \rangle$$
$$\uparrow$$
$$m^{th} \text{ memory location value}$$

Set the Value at Memory Location m to Zero

$$M_m = (b_m)^{N_m}$$

The above expression for $M_m$ is symbolic. The expression represents the following expression in which the operators are carefully ordered to avoid complications (c-numbers etc.) resulting from reordering.

$$M_m \equiv e^{N_m \ln b_m} = \sum_q \frac{(\ln b_m)^q N_m^q}{q!}$$

where the sum ranges from 0 to ∞. $M_m$ becomes

$$M_m = 1 + (b_m - 1)N_m$$

using the identity $N_m = N_m^2$. When $M_m$ is applied to a state it sets the value of the $m^{th}$ memory location to zero.

$$M_m | \ldots, x, \ldots \rangle = | \ldots, 0, \ldots \rangle$$
$$\uparrow$$
$$m^{th} \text{ memory location value}$$

Change the Value at Memory Location m from 0 to the Value at Location n

$$P_m^n = (b_m^\dagger)^{N_n}$$

The above expression for $P_m^n$ is also symbolic. The expression represents the following expression in which the operators are carefully ordered to avoid complications (c-numbers etc.) resulting from reordering.

$$P_m^n \equiv \sum_q \frac{(\ln b_m^\dagger)^q N_n^q}{q!}$$

where the sum over q ranges from 0 to ∞. Using the identity $N_m = N_m^2$ the expression for $P_m^n$ simplifies to:

$$P_m^n = 1 + (b_m - 1)N_m$$

When $P_m^n$ is applied to a state it changes the value of the $m^{th}$ memory location from zero to the value at the $n^{th}$ memory location.

$$m^{th} \quad n^{th}$$
$$\downarrow \quad \downarrow$$
$$P_m^n | \ldots, 0, \ldots, x, \ldots \rangle = (b_m^\dagger)^x | \ldots, 0, \ldots, x, \ldots \rangle$$
$$= | \ldots, x, \ldots, x, \ldots \rangle$$

We can use the operators $M_m$ and $P_m^n$ to express bit-wise assembly language instructions:

**LOAD m** – load the value at memory location m into the register

$$P_r^m M_r = (1 - N_r + b_r)(1 - N_m) + (N_r + b_r^\dagger)N_m$$

The first term on the right handles the case $N_m = 0$ and the second term on the right handles the case $N_m = 1$.

**STORE m** – store the value in the register at memory location m

$$P_m^r M_m = (1 - N_m + b_m)(1 - N_r) + (N_m + b_m^\dagger)N_r$$

The first term on the right handles the case $N_r = 0$ and the second term on the right handles the case $N_r = 1$.

**ADD m** – add the value at memory location m to the value in the register

$$(b_r^\dagger)^{N_m} = \sum_q \frac{(\ln b_r^\dagger)^q N_m^q}{q!}$$

$$= 1 + (b_r^\dagger - 1)N_m$$

If both the register and memory bit m have values of one then the application of this operator expression to the quantum state produces zero.

**SUBTRACT m** – subtract the value at memory location m from the value in the register

$$(b_r)^{N_m} = \sum_q \frac{(\ln b_r)^q N_m^q}{q!}$$

$$= 1 + (b_r - 1)N_m$$

If the value in the register is zero and the value at location m is one the application of this operator produces zero.

**MULTIPLY m** – multiply the value in the register by the value at memory location m

$$(b_r^\dagger)^{(N_m - 1)N_r} = \sum_q \frac{(\ln b_r^\dagger)^q (N_r)^q (N_m - 1)^q}{q!}$$

$$= 1 + (b_r - N_r)(1 - N_m)$$

Other assembly language instructions can be expressed in algebraic form as well.

The operator algebra that we have developed for a bit-wise Quantum Assembly Language™ or a Quantum Machine Language™ provides a framework for the investigation of the properties of Quantum Languages within an algebraic framework – a far simpler task than the standard quantum linguistic approaches.

**Quantum High Level Computer Language Programs**

The Quantum Assembly Language™ representation that we have developed earlier in this chapter forms a basis for high level Quantum Programming Languages. These languages are analogous to high level computer languages such as C or C++ or FORTRAN.

In ordinary computation a statement in a high level language such as

$$a = b + c;$$

in C programming is mapped to a set of assembly language by a C compiler. A simple mapping of the above C statement to assembly language would be

        LOAD ab
        ADD  ac
        STORE aa

where aa is the memory address of a, ab is the memory address of b and ac is the memory address of c.

If we decide to define a High Level Quantum Computer Language™ then it would be natural to define it analogously in terms of a Quantum Assembly Language™. A statement in the High Level Quantum Computer Language™ would map to a set of Quantum Assembly Language™ instructions.

For example, a = b + c would map to the algebraic expression

$$P_{aa}^{r} M_{aa} (a_r^\dagger)^{N_{ac}} \frac{\sqrt{N_r!}}{\sqrt{(N_r + N_{ac})!}} P_r^{ab} M_r$$

using the formalism developed earlier in this chapter to LOAD, ADD and STORE.

The definition of high level Quantum Computer Languages™ in this approach is straightforward. One can then imagine creating programs in these languages for execution on Quantum Computers just as ordinary programs are created for ordinary computers.

Another approach to higher level Quantum Computer Languages™ is to simply express them directly using raising and lowering operators – not in terms of Quantum Assembly Language™ instructions. For example the preceding a = b + c; statement can be directly expressed as

$$(a_{aa}^\dagger)^{N_{ac}+N_{ab}} (a_{aa})^{N_{aa}} \frac{1}{\sqrt{(N_{ac} + N_{ab})!} \sqrt{N_{aa}!}}$$

Simple High Level Quantum Computer programs can be expressed as products of algebraic expressions embodying the statements of the program. These programs are sharp on the set of memory states taking an initial memory state that is an eigenstate of the set of number operators $N_m$ into an output eigenstate of the number operators.

A general High Level Quantum Computer Program is a sum of simple High Level Programs. For example,

$$\alpha h_1() + \beta h_2() + \gamma h_3() | \ldots >$$

where $\alpha, \beta,$ and $\gamma$ are constants such that

$$|\alpha|^2 + |\beta|^2 + |\gamma|^2 = 1$$

The sum of simple programs $\alpha h_1() + \beta h_2() + \gamma h_3()$ produces the state $| n_1, m_1, p_1, \ldots >$ with probability $|\alpha|^2$, the state $| n_2, m_2, p_2, \ldots >$ with probability $|\beta|^2$, and the state $| n_3, m_3, p_3, \ldots >$ with probability $|\gamma|^2$.

An initial eigenstate of the number operators is tranformed into an output state that is a superposition of number operator eigenstates. In this case we use probabilities to specify the likelihood that a given output eigenstate will be found when the output state is measured.

A Hamiltonian can also be used to specify the time evolution of a system starting from an initial state. Using the notation:

$$| \Psi(t) >$$

to denote the state of a Quantum Computer at time t the general state of a computer at time t can be written as a superposition of number representation states:

$$| \Psi(t) > = \sum_n f_n(t) | n_1, n_2, n_3, \ldots >$$

where n represents a set of values $n_1, n_2, n_3, \ldots$

The time evolution of the states can be specified using the Hamiltonian operator H as

$$|\Psi(t)\rangle = e^{-iHt}|\Psi(0)\rangle$$

A simple example of a Hamiltonian that causes a Quantum Computer to evolve in a non-trivial way is:

$$H = \sum_{m=0}^{\infty} (a_{m+2}^{\dagger})^{N_{m+1}+N_m} (a_{m+2})^{N_{m+2}} \frac{1}{\sqrt{(N_{m+1}+N_m)!} \sqrt{N_{m+2}!}}$$

This Hamiltonian is based on the a = b + c statement above. This Hamiltonian generates a complex superposition of states as time evolves. More complex Hamiltonians equivalent to programs with several statements can be easily constructed.

### Quantum C Language™

One of the most important computer languages is the C programming language developed at Bell Laboratories in the 1970's. The original version of version of the C language was a remarkable combination of low level (assembly language-like) features and high level features like the mathematical parts of FORTRAN. The variables in the language were integers stored in words just as we saw in the earlier examples in this chapter. (There were several other types of integers as well – a complication that we will ignore.)

Using the ideas seen in the earlier sections of this chapter it is easy to develop algebraic equivalents for most of the constructs of the C language and thus create a Quantum C Language™. An important element that must be added to the previous development is to introduce the equivalent of pointers. Simply put pointers are variables that have the addresses of memory locations as their values. The C language has two important operators for pointer manipulations:

| Operator | Role | Example |
| --- | --- | --- |
| & | Fetch an address | ptr = &x; |
| * | Fetch/set the value at an address | z = *ptr;<br>*ptr = 99; |

The & operator of C fetches the address of a variable in memory. The example shows a pointer variable ptr being set equal to the address of the x variable. The * (dereferencing) operator can fetch the value at a memory location. The first * example illustrates this aspect: the variable z is set equal to the value at the memory location specified by the pointer variable ptr. The * operator can also be used to set the value at a memory location as illustrated by the second * example. In this example the value 99 is placed at the memory location (address) specified by the ptr pointer variable.

These operators can be implemented in the algebraic representation of the Quantum C Language™ in the following way:

$$\& \quad \Leftrightarrow \quad [A, \ ]$$

where $A = \Sigma\, m(a_m - a_m^{\dagger})$ with the sum from 0 to $\infty$. If we apply the operator to a raising or lowering operator we obtain its address

$$\&\, a_m^{\dagger} = [A, a_m^{\dagger}] = m = \&\, a_m = [A, a_m]$$

The equivalent of the * operator is actually a pair of operator expressions. To fetch the value at a memory location we use

$$*m \equiv N_m$$

To set the value to X at a memory location m we use a more complex C language representation:

$$*m = X;$$

An equivalent algebraic expression is:

$$*m(X) \equiv (a_m^\dagger)^X (a_m)^{N_m} \frac{1}{\sqrt{X!} \sqrt{N_m!}}$$

*m(X) is a functional notation. So a = b + c can be rewritten as a "pointer" algebraic expression as:

$$(a_{aa}^\dagger)^{*ac + *ab} (a_{aa})^{*aa} \frac{1}{\sqrt{(*ac + *ab)!} \sqrt{*aa!}}$$

Or more compactly using the functional notation as

$$*aa(*ab + *ac)$$

The Quantum C Language™ could be used to define Hamiltonians for a Quantum Computer. Other languages such as Java™, C++, lisp and so on also have Quantum analogues which may be defined in a similar way.